# Ionic effect on combing of single DNA molecules and observation of their force-induced melting by fluorescence microscopy

Yu-Ying Liu, Peng-Ye Wang[*], Shuo-Xing Dou, Wei-Chi Wang, Ping Xie, Hua-Wei Yin, Xing-Dong Zhang

*Laboratory of Soft Matter Physics, Institute of Physics,*
*Chinese Academy of Sciences, P.O. Box 603, Beijing 100080, China*

**Abstract**

Molecular combing is a powerful and simple method for aligning DNA molecules onto a surface. Using this technique combined with fluorescence microscopy, we observed that the length of lambda-DNA molecules was extended to about 1.6 times their contour length (unextended length, 16.2 micrometers) by the combing method on hydrophobic polymethylmetacrylate (PMMA) coated surfaces. The effects of sodium and magnesium ions and pH of the DNA solution were investigated. Interestingly, we observed force-induced melting of single DNA molecules.

**1. Introduction**

The study of DNA is being greatly advanced by the fluorescence microscopy technique at the single molecule level [1, 2]. Recently, DNA stretching has attracted much attention. Several physical methods have been employed to stretch single DNA molecules, such as magnetic tweezer [3–5], laser tweezer [6–9], micropipette [10] and fluid flow. Up to now, various fluid-flow induced DNA fixation and elongation techniques have been reported, e.g., receding of an air-water interface [11–13], gas flow driven droplet motion [14], sliding of a coverslip edge [15], convective fluid flow [16], spin-stretching [17], precisely controlled meniscus motion [18]. In this work, we investigated the molecular combing of DNA by evaporation of small droplets of DNA solution on hydrophobic polymethylmetacrylate (PMMA) surfaces. This type of surfaces was previously used for studying molecular combing [19] and for observation of transcription activity of RNA polymerase on single combed DNA molecules [20]. In our experiments we studied the effects of sodium ion, magnesium ion and buffer pH on molecular combing of DNA on PMMA surfaces and, more interestingly, we observed force-induced melting of single DNA molecules.

**2. Materials and methods**

*2.1. Surface treatment of glasses*

Glasses (25.4 x 76.2 mm, 1-mm thickness; China) were immersed in hot NaOH (1 M) for about 1 hour, and then rinsed thoroughly in high-purity water (Millipore S.A., France). Glass surfaces were rendered hydrophobic by coating the surfaces with PMMA by using the same method as that used in Ref. [20]. Briefly, a droplet (0.2 ml) of PMMA (560F, Japan) in chloroform [10% (wt/wt)] was dripped down onto the center of a cleaned glass, which was mounted horizontally on the spin-coating machine using a double-sided tape, and spread by spin-coating at 2,500 rpm for 1 min. After spin-coating, the PMMA film was formed evenly on the whole glass surface. To further improve the film uniformity, the glass was then baked at 165 °C for ~ 20 min and then stored at room temperature in a dust-free environment. During the preparation of PMMA film, it was very important to keep the uniformity of the PMMA solution and the cleanness of the glasses.

---
[*] Corresponding author. E-mail address: pywang@aphy.iphy.ac.cn

*2.2. DNA preparations*

We used Lambda DNA (Sino-American Biotechnology Company, China) in our experiments. Lambda DNA (48.5 kb) was stained with a fluorescent dye, oxazole yellow dimmer (YOYO-1, Molecular Probes) at a ratio of ten base pairs per dye molecule (bp/dye = 10) by mixing DNA sample with a specific volume of freshly prepared 0.1-μM dye solution (10 mM Tris /1 mM EDTA buffer, pH 8.0). The DNA/YOYO-1 solution was incubated for approximately 30 min in a dark room, and then was diluted to 6.5 pM in a 50 mM Bis-Tris buffer (pH 6.6, Sigma).

*2.3. Molecular combing*

A droplet (1.2 μl) of the stained DNA solution was deposited onto a PMMA surface. At first, the interface of droplet didn't move at all. After about 5~10 minutes, the air-water interface started to recede due to significant drying of the droplet, and DNA molecules originally bound to the surface with one extremity were extended and immobilized on the dried surface, whereas unbound molecules were swept away by the moving interface. As a result, a significant number of fixed DNA molecules were fully elongated and aligned radially. As most of them were concentrated near the droplet's round periphery, usually a "sunburst" pattern was formed.

*2.4. DNA imaging by fluorescence microscopy*

Fluorescently stained DNA molecules were observed by using an inverted optical microscope (IX-70; Olympus) by epifluorescence with a 20μ objective. The images were captured by a cooled CCD camera (CoolSNAP-fx, 1300μ1030 pixels, 12-bit digitization; Roper Scientific, Inc.). YOYO-1 has an excitation maximum at 491 nm and emission maximum at 509 nm. A 100 W mercury lamp was used in combination with a U-MWB excitation cube (BP450-480, DM500, BA515). The CCD acquisition time we used was three seconds. The lengths of the elongated molecules were measured by using a Scion image software.
To reduce the illumination time and thus the fluorescence photobleaching of YOYO-1, we generally started our fluorescence microscopy observation after the droplets dried completely. In addition, this also avoided photocleavage of DNA in the solution [21]. It should be noted that although both photobleaching and photocleavage were expected to occur during observation of the combed molecules, photocleavage should have no visible effect on the observed molecules because even if a DNA was cleaved, the two new ends should remain close together on the dry surface. This is in contrast with the case when photocleavage occurs to stained DNA in solution, where a single fluorescent DNA molecule may be broken up into some smaller fragments [21].

### 3. Results and discussion

Many factors affect the combing procedure, such as surface type, DNA concentration and solution pH, ion strength, temperature, and droplet size. The appropriate moving velocity of the air-water interface is also important in the forming of well combed DNA [18].

*3.1. Hydrophobic surface characteristics*

Surface characteristics have a great effect on the formation of combed DNA. Excessively strong adsorption prevents molecular elongation, whereas weak adsorption does not fix a sufficient number of molecules to the surface [15, 19]. For comparison, we also prepared polystyrene (PS) surfaces in our experiments. PS surfaces have a stronger hydrophobic nature than PMMA. We found that the air-water interfaces moved much slower on PS surfaces than on PMMA surfaces and DNA cannot be well combed on PS surfaces (data not shown) with our present experimental conditions.

On PMMA surfaces, well-combed single DNA molecules were observed. One typical result is given in Fig. 1(a), where part of the original droplet periphery can be seen clearly as represented by a well defined boundary. From the corresponding extension length distribution of the combed DNA (see Fig. 2(d)) it can be seen that most of the combed molecules have extension lengths around 27 μm, i.e., ~ 1.6 times their contour B-form length (~ 16.2 μm). It should be noted that Bensimon and coworkers [11, 12, 19] have obtained extension lengths of the same molecules ranging from 16 μm to 27 μm on different types of surfaces, with the 27-μm extension obtained by combing on hydrophobic polystyrene surfaces.

It was when the air-water interface was receding that the force was exerted on DNA molecules, so the force was related to the air-water interface tension [11]. During the evaporation of a droplet, there should be some variation of the velocity of the receding interface. We found this variation has no visible effect on the extension length of DNA molecules.

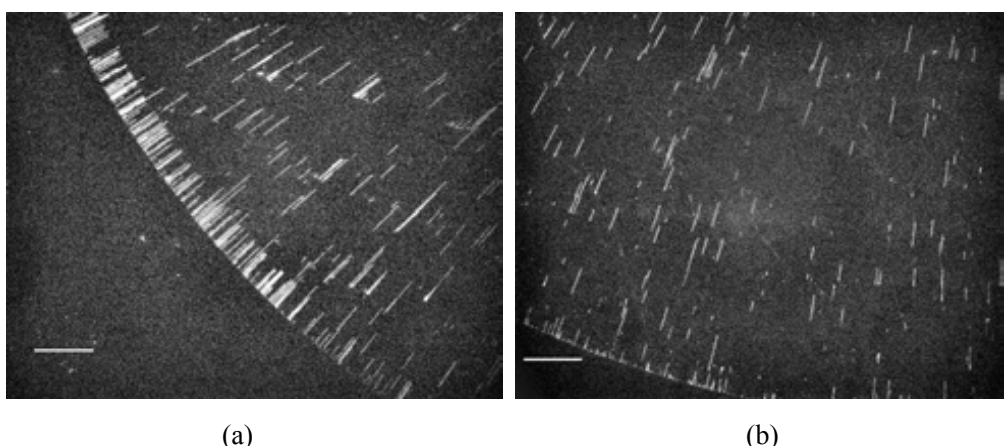

(a)                                         (b)

Fig. 1. CCD images of single combed DNA molecules with buffers of 6.5 pM DNA at pH 6.6, and different $Na^+$ concentrations: (a) 150 mM; (b) 25 mM. [Bars: 50 μm (a, b).]

*3.2. Ionic effect*

The effect of sodium and magnesium ions on the elongation of DNA molecules was investigated. We found that both metal ions enhanced DNA binding to PMMA surfaces. This should be due to the fact that these ions increase the hydrophobic interaction between DNA and the PMMA surface by increasing the molal surface tension of water. This ionic effect has long been known and used in promoting ligand-protein interactions and thus protein precipitation in hydrophobic interaction chromatography (HIC) [22]. On the other hand, under the condition of higher ionic strength, the fluorescence of the stained DNA was much dimmer. We believe this should be caused by reduction of the binding efficiency of YOYO-1 in the presence of the ions. It is known that when the mixing ratio is below 0.125 dye molecule per base pair, the predominant binding mode of YOYO-1 on DNA is bisintercalation; when the mixing ratio is above 0.125, groove association (external binding) with DNA begins to contribute significantly [23]. The relevant experimental condition is that the mixing solution was equilibrated for more than 48 h. In this case, an increase in the contour length of DNA should be observed.

In fact, due to the presence of four positive charges on the molecule, YOYO-1 is characterized by very high electrostatic binding affinity for DNA [21]. When the time allowed for the mixing solution to equilibrate was not so long, it was observed that YOYO-1 bind to DNA without significant effect on the intrinsic contour length of the molecule, implying that electrostatic interactions and groove associations are mainly involved in the binding [21]. In addition, Bennink et al. [9] had studied the interaction of DNA with YOYO-1 by measuring the force-extension curves of single DNA molecules in a buffer with YOYO-1 (DNA was incubated for 15 min with YOYO-1 before force was applied). They observed, depending on the stretching speed, a 10-20% extension of the DNA molecule. When stretching with high speed (7.1 seconds for a whole stretch-relax cycle), the force-extension curve was no different from that obtained by other authors without YOYO-1, only the force-extension curve during the relax phase showed an extension of its intrinsic length by ~10%. We believe the original binding mode of YOYO-1 was external binding, and the extension was caused by YOYO-1 intercalation during stretching of the molecule. This explains why the longer the stretch-relax time, the larger the extension of the molecule was. From their experiments we may also know that the YOYO-1 molecules abandoned gradually their intercalation binding during the relax phase.

As our DNA/YOYO-1 solution was incubated for only about 30 min, we think the dominant binding mode of YOYO-1 should also be external binding in our case. That is, the binding is mainly of electrostatic nature. Thus the positively charged Na and Mg ions should reduce the binding efficiency of the positively charged YOYO-1 molecules by competitive binding to DNA. We found that even if YOYO-1 and DNA were mixed first and NaCl or $MgCl_2$ was added into the solution some time later, the fluorescence intensity of the stained DNA was still significantly reduced.

In the following two subsections, we mainly give the results about the effect of sodium and magnesium ions on the extension length of combed DNA. It was observed that at low ionic strength, usually a large amount of short fragments existed among the combed molecules.

*3.2.1. Sodium ion*

In the presence of Na$^+$, we found that many DNA molecules were broken, especially when the ionic strength is low (< 100 mM). A typical result is given in Fig. 1(b). As DNA was destabilized at low ionic strength, the fragmentation of DNA should occur both before combing (i.e., in solution) and during combing.

To obtain reliable and systematic results for the effect of Na$^+$ on the extension of DNA, we had used Na$^+$ concentrations in a large range (0-200 mM) under otherwise identical conditions: at room temperature, on the same surface, using the same droplet volume (1.2 μl). Typical results on the length distribution of combed DNAs are given in Fig. 2. We found the maximal position of relatively high columns was still around 27 μm in most cases and obviously a relatively large amount of fragments exist with [Na$^+$] between 5 mM and 100 mM.

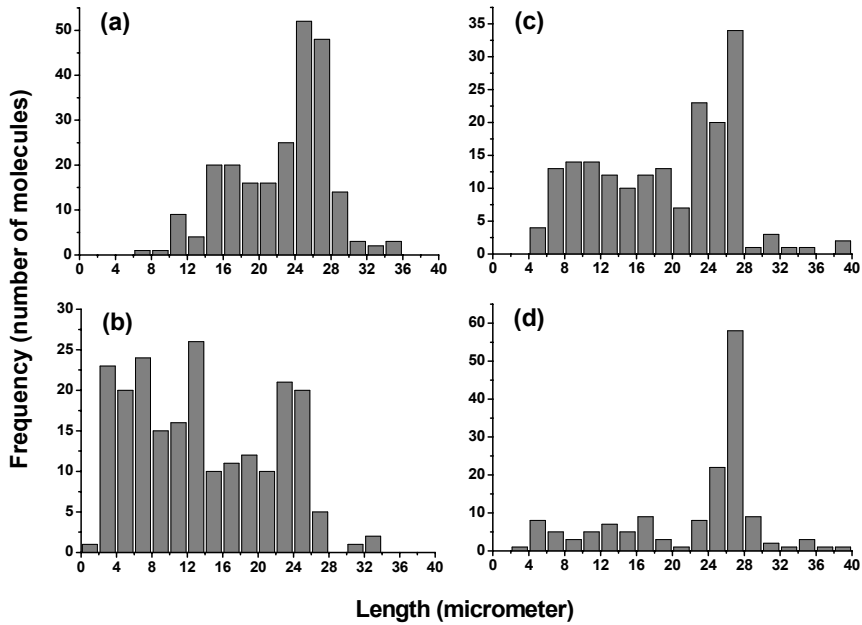

Fig. 2. Histograms of the DNA length with buffers of 6.5 pM DNA at pH 6.6 and different [Na$^+$]: (a) 0 mM, $N$ = 235; (b) 25 mM, $N$ = 220; (c) 50 mM, $N$ = 187; (d) 150 mM, $N$ = 157.

To show quantitatively the results about the ionic effect on the fragmentation, variations with [Na$^+$] of the mean length of combed DNA and the corresponding percentage of short molecules with extension lengths less than 24 μm are presented in Fig. 3, where lower mean length and higher percentage mean more fragmentation. It can be seen that the fragmentation was most serious at 25 mM [Na$^+$]. At 0 mM [Na$^+$], on the contrary, the fragmentation was not very different from that at high [Na$^+$]. It should be noted that Clausen-Schaumann et al. [24] had studied the stability of single DNA molecules by measuring the force-extension curves under different buffer conditions. The behavior of DNA in a buffer without Na$^+$ was similar to that of single-stranded DNA, implying that large part of the molecule melts as soon as it was stretched. But the molecules could be stretched by a force of more than 250 pN with rupture. This is consistent with our results.

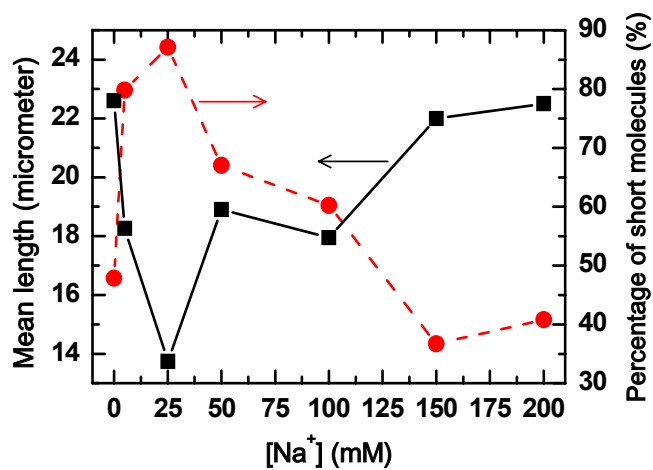

Fig. 3. [Na$^+$] dependences of the mean length of combed DNA molecules and the percentage of short ones (the extension length < 24 μm). The buffer condition was 6.5 pM DNA at pH 6.6.

*3.2.2. Magnesium ion*

The effect of Mg$^{2+}$ on the combing of DNA was stronger than that of Na$^+$ at a given concentration. When [Mg$^{2+}$] was as low as 0.5 mM, although the length of combed DNA molecules and the fluorescence intensity were almost the same as that with no Mg$^{2+}$, the binding efficiency was significantly improved. Results with 1 mM [Mg$^{2+}$] are given in Fig. 4 together with that of no Mg$^{2+}$ for comparison. As in the case of Na$^+$, with the increase of [Mg$^{2+}$] from 0 to 10 mM, the number of short DNAs increased over the same range. As [Mg$^{2+}$] was higher than 10 mM, the fluorescence became too weak for clear observation. Thus the effect of [Mg$^{2+}$] on the fluorescence intensity was much more serious than Na$^+$. This implies that the divalent Mg ions reduce more seriously the DNA/YOYO-1 interaction by competitive binding to DNA than the monovalent Na ions. In fact, Mg$^{2+}$ had been observed to reduce the persistence length of DNA much more effectively than Na$^+$ [25].

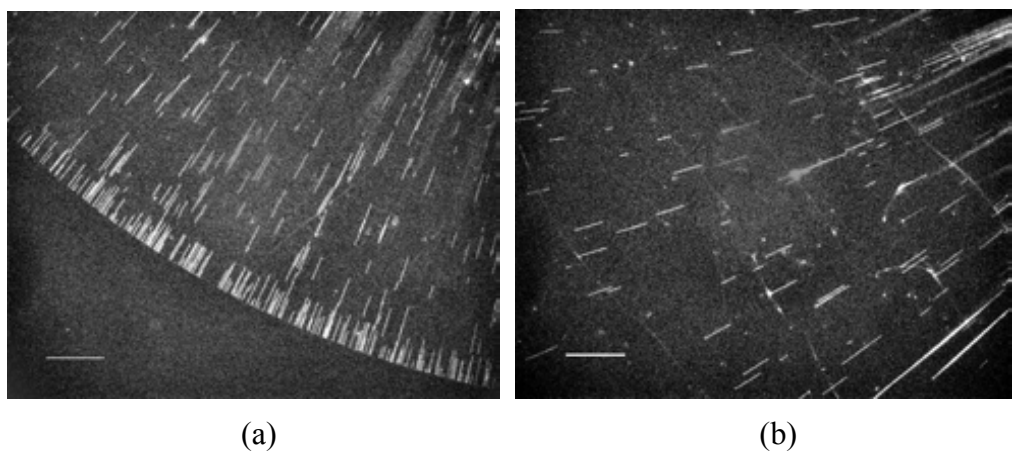

(a)          (b)

Fig. 4. CCD images of single combed DNAs with buffers of 6.5 pM DNA at pH 7.5 on the same surface: (a) 1 mM Mg$^{2+}$, (b) no Mg$^{2+}$. [Bars: 50 μm (a, b).]

Typical results on the length distribution of the combed molecules are given in Fig. 5. Note the difference between Fig. 2(a) and Fig. 5(a). As the ionic strength was essentially zero in both cases, the slight difference should be caused by a difference in the buffer pH.

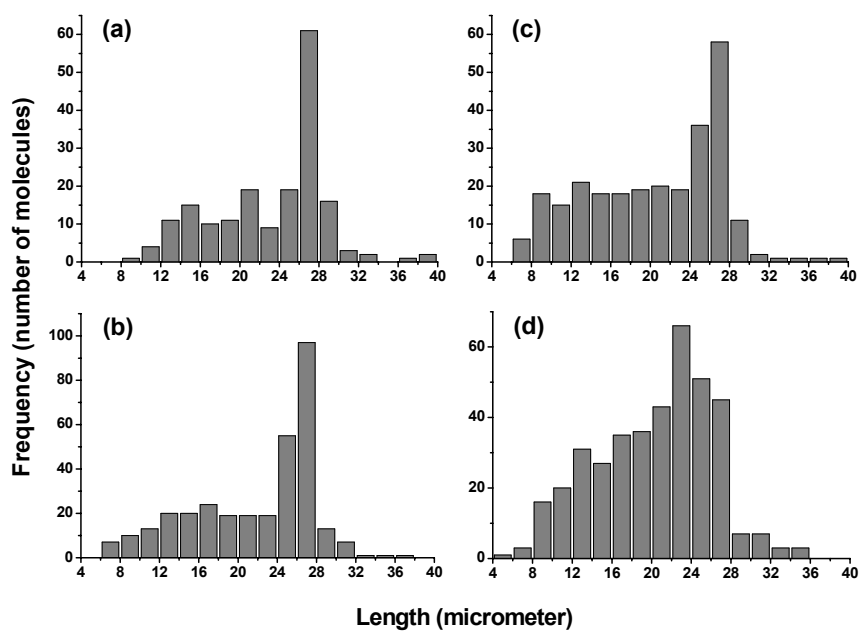

Fig. 5. Histograms of the DNA length with buffers of 6.5 pM DNA at pH 7.5 and different [$Mg^{2+}$]: (a) 0 mM, $N$ = 185; (b) 1 mM, $N$ = 326; (c) 5 mM, $N$ = 265; (d) 10 mM, $N$ = 396.

The variation of mean length of combed DNA with [$Mg^{2+}$] is presented in Fig. 6 together with the percentage of short molecules with extension lengths less than 24 μm. From the results for $Na^+$ (Fig. 3) we expect that the mean length would rise and the percentage drop at high [$Mg^{2+}$].

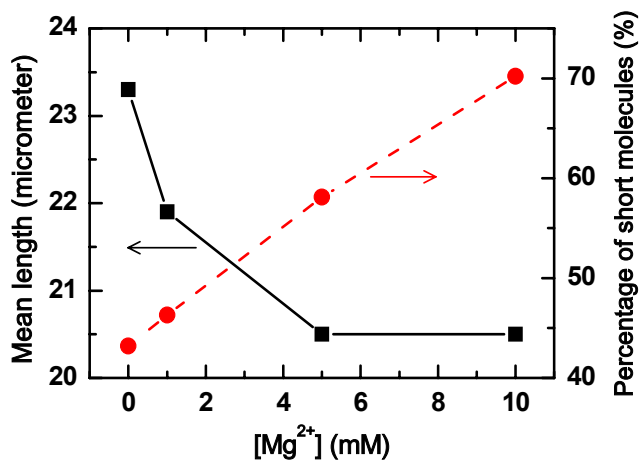

Fig. 6. [$Mg^{2+}$] dependences of the mean length of combed DNA molecules and the percentage of short ones (the extension length < 24 μm). The buffer condition was 6.5 pM DNA at pH 7.5.

*3.3. Effect of buffer pH*

The effect of buffer pH on specific binding and combing of DNA on a great variety of surfaces including PMMA has been well characterized and explained by Allemand et al. [19]. They demonstrated that specific binding of double-stranded DNA via its unmodified extremities can be achieved on various surfaces at appropriate pH. Extreme of low pH results in strongly and nonspecifically bound molecules that cannot be elongated by a receding air-water interface. Extreme of high pH, on the other hand, results in weakly bound and extended molecules. For the case of neutral hydrophobic surfaces like PMMA, the reason is that, at low pH, the DNA bases undergo intensive protonation that induces partial DNA melting. This exposes the hydrophobic core of the helix and, as a result, the affinity between the DNA molecule and the hydroscopic surface is greatly enhanced. At high pH, the affinity is low because DNA is repelled by its image charge [19].

Our experimental results with PMMA confirmed the above observations. At low pH, the molecules appeared as bright spots, only a few were combed. At medium pH, many molecules were combed. At high pH, a few molecules were adsorbed and combed. Typical results are given in Fig. 7. We found that the appropriate pH for good combing of DNA was between 5 and 7, in agreement with that in Ref. [19].

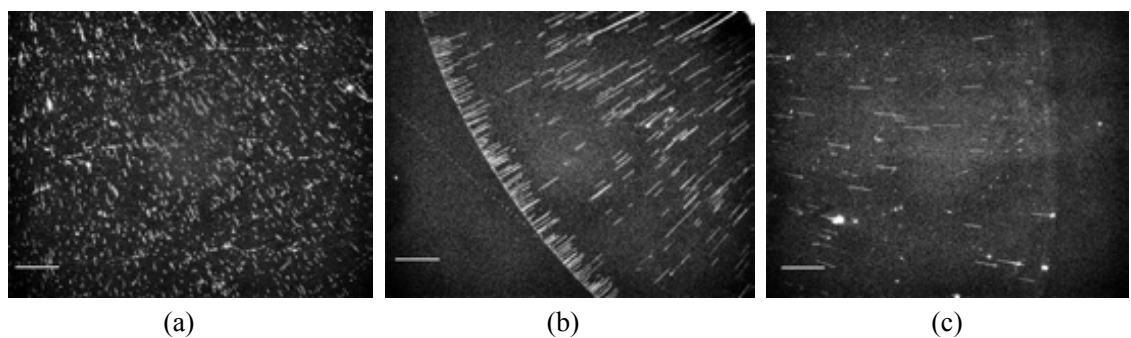

(a)  (b)  (c)

Fig. 7. CCD images of the adsorbed and combed DNA molecules on the same surface with different buffer pH (no NaCl added). (a) pH 3; (b) pH 6.6; (c) pH 12. At this high pH, many molecules may have been melted in the solution, but single-stranded DNAs could not be observed when adsorbed on the surface. [Bars: 50 μm (a–c).]

*3.4. Force-induced melting*

DNA is an extensible molecule and can be stretched beyond its B-form contour length under high force, and a remarkable discontinuity in the elastic response of double-stranded DNA occurs at ~ 60-70 pN. At this force, the molecule abruptly, but reversibly, assumes an extended form about 1.7 times of its B-form contour length. The force rises rapidly again as the extension becomes greater [8, 10]. That is, there exists a force plateau from its fully extended B-form length to 1.7 times its B-form length in the force-extension curves. Actually, the height of the plateau varies with buffer conditions such as pH and salt concentration [8, 26–28]. Such a plateau signifies a highly cooperative overstretching transition from B-form to S-form DNA.

As to the S-form structure, it was proposed that the overstretched DNA adopts a diameter-reduced double-stranded structure [10], or ladder-like structure [8]. Recently it was proposed that during the overstretching transition, DNA melting occurs [26–30]. And as domains of melted DNA are separated by helical sections, both DNA strands still remain close together during the transition. At the end of the transition, few short helical domain boundaries hold the largely melted strands together [27, 28].

In our experiment, under normal buffer conditions (pH 6.6, [$Na^+$] > 100 mM), the extension lengths of combed DNA molecules usually were focused on ~ 1.6 times of their contour length. Referring to force-extension curves obtained by single-molecular manipulation under similar buffer conditions [28], we know that most combed molecules were in overstretched forms achieved with forces around 65 pN. Among the different structures mentioned above, whatever structure the overstretched DNA adopts, YOYO-1 cannot bind to DNA by intercalation, especially near the end of the overstretching transition. The fact that YOYO-1 still remained bound to the overstretched DNA confirms that the binding mode of YOYO-1 was indeed external binding. In fact, this should be also true for the case of Bensimon et al. [12], where during combing, stained DNA had been stretched to 2.14 times its unstretched length before broken; however, YOYO-1 still remained bound to the molecule.

As the forces acting on the DNA molecules during combing were ~ 65 pN, forced-induced melting of DNA might occur. But if the melted strands remain close together as proposed in Refs. [27, 28], YOYO-1 may still binds to them, thus the melting cannot be observed. However, if one of the melted strands is nicked and frays back from both sides of the nick [8], we expect that YOYO-1 will be absent in that melting region as YOYO-1 does not bind to single-stranded DNA.

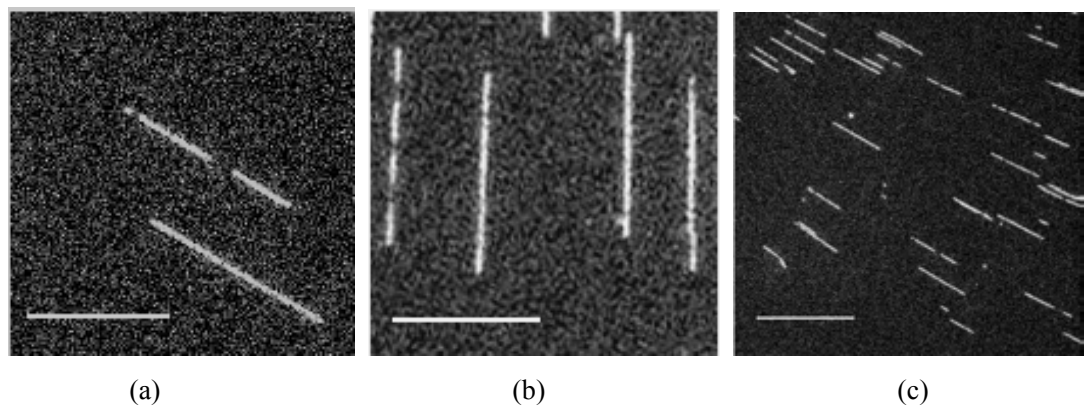

(a)  (b)  (c)

Fig. 8. Selected images showing DNA melting (150 mM $Na^+$). [ Bars: 20 μm (a, b); 50 μm (c).]

Interestingly, we had indeed observed DNA molecules with one or more dark sections interlacing with bright sections (see Fig. 8). We believe this phenomenon corresponds to melting rather than rupture of DNA for the following several reasons. (i) If the dark sections were caused by rupture of DNA during combing, the total length of the bright and dark sections would be larger than the extension length of a single intact molecule (by a value corresponding the total length of the dark sections). In our experiments, however, the two lengths were similar (also see the following Fig. 9). (ii) Further, if a DNA was broken once or sequentially several times during combing, the chance should be rare that the falling fragments immediately bind to the surface nearby their respective break sites. We believe a falling fragment would rather be swept away by the fast receding air-water interface. This can be seen from the fact that in the case of low salt concentration, DNA should be easily broken during combing, but the above phenomenon was not so frequent. Rather, it was at high salt concentration that the melting phenomenon was more frequent. (iii) If photocleavage occurs during our observation, a dark region cannot be formed because the whole molecule is immobilized on the dry surface. In addition, the appearance of dark sections was not equally frequent in cases of different original buffer conditions, and no new dark section was found to appear during our observation.

We found force-induced melting phenomenon could be observed with different DNA concentrations (e.g., 2 pM, 3.25 pM and 6.5 pM) and $Na^+$ concentrations (from 0 to 200 mM). But it was observable more easily at 150 mM $Na^+$ in our present experiments. The reason may be: the more stable the DNA molecule is, the higher the possibility that a single DNA strand remains intact after its complementary strand is ruptured. We measured the total lengths of melted DNA molecules. The result is given in Fig. 9. It can be seen that the peak value was around 28 μm, only slightly larger than lengths of intact molecules (~ 27 μm).

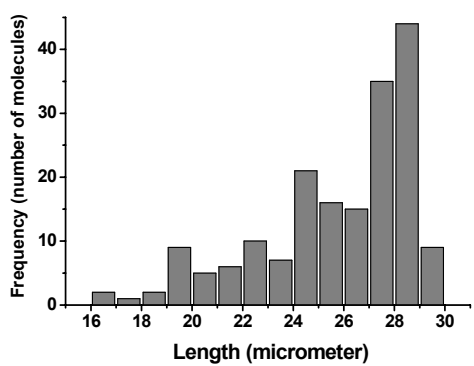

Fig. 9. Histogram of extension length of melted DNA molecules (6.5 pM DNA, pH 6.6, [$Na^+$] from 0 to 200 mM, $N = 184$).

## Conclusions

In summary, we studied molecular combing of single lambda-DNA on PMMA surfaces under different Na and Mg ionic and pH conditions. Both Na and Mg ions enhanced DNA binding to PMMA surfaces while low ionic strengths of both $Na^+$ and $Mg^{2+}$ caused a large amount of short fragments in the combed molecules. In the case of $Na^+$, the fragmentation was most serious at [$Na^+$] = 25 mM. At high ionic strengths, the lengths of the combed molecules focus on ~ 27 μm, ~ 1.6 times of their contour B-form length. Both ions reduced, in proportion to their concentrations, the fluorescence intensity of stained molecules, and the effect of $Mg^{2+}$ was much more serious than that of $Na^+$. The buffer pH appropriate for molecular combing on PMMA surfaces is between 5 and 7.

Combing force-induced melting of DNA molecules was observed under different buffer conditions. Usually one to three melting sections might be observed on a single molecule. The melting phenomenon could be most easily observed at the concentration of 150 mM $Na^+$ in our present experiments. The binding mode of YOYO-1 to DNA was believed to be external binding in our experiments. Thus YOYO-1 remained binding to DNA when it was overstretched. We think that in a melting section, one of the two DNA strands was nicked and frayed back from both sides of the nick. But it might also be possible that both strands were intact, but not close enough to each other for the binding of YOYO-1. We expected this could be determined by combing with other methods such as atomic force microscopy.

**Acknowledgements**

The project is supported by the National Natural Science Foundation of China (Grant Number 60025516), and the Innovation project of the Chinese Academy of Sciences.